\documentclass[12pt]{iopart}

\begin{document}

\title{Superconductivity induced by Ni doping in BaFe$_2$As$_2$ single crystals}

\author{L J Li, Y K Luo, Q B Wang, H Chen, Z Ren, Q Tao, Y K Li, X Lin, M He, Z W Zhu, G H Cao\footnote[1]{Electronic address: ghcao@zju.edu.cn}, Z A Xu\footnote[2]{Electronic address: zhuan@zju.edu.cn}}

\address{Department of Physics, Zhejiang University, Hangzhou 310027, People's Republic of China}

\begin{abstract}
A series of 122 phase BaFe$_{2-x}$Ni$_x$As$_2$ ($x$ = 0, 0.055,
0.096, 0.18, 0.23) single crystals were grown by self flux method
and a dome-like Ni doping dependence of superconducting transition
temperature is discovered. The transition temperature $T_c^{on}$
reaches a maximum of 20.5 K at $x$ = 0.096, and it drops to below
4 K as $x$ $\geq$ 0.23. The negative thermopower in the normal
state indicates that electron-like charge carrier indeed dominates
in this system. This Ni-doped system provides another example of
superconductivity induced by electron doping in the 122 phase.

\end{abstract}


\maketitle

\section{Introduction}
Following the discovery of superconductivity at $T_c$ of 26 K in
F-doped LaFeAsO\cite{Hosono}, a so-called 1111 phase family of
high-$T_c$ superconductors with ZrCuSiAs-type structure, $i.e.$,
LnFeAsO$_{1-x}$F$_{x}$\cite{WNL-LaOF,WHH-LaOF,Sefat-LaOF,WNL-CeOF,Chen-SmOF,ZZX-PrOF,ZZX-NdOF},
LnFeAsO$_{1-\delta}$\cite{ZZX-OH} and
Ln$_{1-x}$Th$_{x}$FeAsO\cite{ZAXu-GdTh,ZAXu-TbTh} (Ln = La, Ce,
Pr, Nd, Sm, Gd, Tb) have been reported. Similar to the cuprate
high temperature superconductors, the (Fe$_{2}$As$_{2}$)-layers
are conducting layers and essential to the occurrence of
superconductivity, while the (R$_{2}$O$_{2}$)-layers inject charge
carriers to the former via chemical doping and also retains
structural integrity of the (Fe$_{2}$As$_{2}$)-layers.
Furthermore, a 122 phase family of high-$T_{c}$ superconductors,
AFe$_{2}$As$_{2}$ (A = Ca, Sr and
Ba)\cite{Johrendt-BaK,LuoJL-SrK,ChuCW-KSr,Chen-BaKBaLa,Chen-CaNa}
with the ThCr$_{2}$Si$_{2}$-type structure (space group I4/mmm),
which possess the same (Fe$_{2}$As$_{2}$)-layers as LnFeAsO but
separated by simple A-layers, have been discovered.
Superconductivity in a so-called 111 phase compound LiFeAs with a
similar structure has also been reported\cite{Pitcher-LiFeAs}. In
the 1111 phase family, superconductivity is mostly induced by
electron doping, i.e., partial substitution of O by F, Ln by Th,
or O-vacancy though superconductivity with $T_{c}$ of 25 K has
indeed been reported in hole-doped
La$_{1-x}$Sr$_{x}$FeAsO\cite{WHH-LaSr}. In contrast,
superconductivity is mostly induced by hole doping in 122 phase,
i.e., partial substitution of Ca, Ba, or Sr by K or Na etc. No
superconductivity has been observed by means of A-site electron
doping, such as the partial substitution of divalent ions
(Ba$^{2+}$ or Sr$^{2+}$) by trivalent ions
(La$^{3+}$)\cite{Chen-BaKBaLa}.

In contrast to high-$T_c$ cuprates, superconductivity can also be
induced by partial substitution of Fe in the conducting layers by
other transition metal elements like Co\cite{Sefat-LaCo,Cao-LaCo}
and Ni\cite{Cao-LaNi} in the 1111 phase, and the measurements of
Hall effect and thermopower indicate that the electron-type charge
carriers dominate in the Co doped case. For the 122 phase,
superconductivity with $T_{c}$ as high as 25 K has also been
observed in Co-doped BaFe$_{2-x}$Co$_{x}$As$_{2}$\cite{BaFeCoAs}
and SrFe$_{2-x}$Co$_{x}$As$_{2}$\cite{SrFeCoAs} systems. As
implied by the negative thermopower value in the normal
state\cite{Chen-BaKBaLa,Cao-LaCo,SmCo}, the Co doping appears to
donate the extra electrons into Fe layers as itinerant charge
carriers, and thus induce superconductivity for either 1111 phase
or 122 phase. In this paper, we report that superconductivity is
induced in the Ni-doped BaFe$_{2-x}$Ni$_{x}$As$_{2}$ single
crystals. A dome-like Ni doping dependence of $T_{c}$ is
established and the highest \emph{T}$_c^{on}$ of 20.5 K is
realized at the optimal doping level $x$ = 0.1. Compared with the
Co doped 122 phase, the superconducting window is narrower and the
optimal Ni doping concentration is only about half. This Ni-doped
system provides another example of superconductivity induced by
electron doping in the 122 phase.

\section{Experimental}
Single crystals with nominal formula BaFe$_{2-x}$Ni$_x$As$_2$ ($x$
= 0, 0.05, 0.1, 0.16, 0.2) were prepared by self flux
method\cite{BaFeCoAs}. All the starting materials, Ba rods, Fe
powders, Ni powders, As pieces, are of high purity($\geq$99.99\%).
First, FeAs and NiAs binaries were prepared by reacting Fe/Ni
powder and As powder in evacuated silicon tube. They were heated
slowly to 873 K, kept for 10 hours, then cooled to room
temperature. A ratio of Ba: FeAs/NiAs = 1:5 was used and the extra
FeAs materials acted as the flux. The mixtures of Ba and FeAs/NiAs
were loaded in a finely designed thin corundum crucibles. These
processes were all done in a glove box filled with high purity
argon. Then the crucibles were sealed in evacuated quartz tubes.
Finally the quartz tubes filled with the starting materials were
slowly heated to 973 K, dwelled for 5 hours and then heated to
1453 K, dwelled for 10 hours. Then the temperature decreased to
1363 K at a rate of -2 K/hour, followed by decanting the quartz
tube at 1363 K, and finally cooled down to room temperature
slowly. Single crystals with a diameter of about 5 mm were
obtained.

X-ray diffraction (XRD) was performed at room temperature using a
D/Max-rA diffractometer with Cu-K$_{\alpha}$ radiation and a
graphite monochromator. Lattice parameters were refined by a
least-squares fit. Chemical analysis by Energy Dispersive of X-ray
(EDX) microanalysis was performed on an EDAX GENESIS 4000 X-Ray
analysis system affiliated to a Scanning Electron Microscope (SEM,
model SIRION). For each doped single crystal, EDX microanalysis
was performed at more than three different spots to check the
distribution of Ni content. The electrical resistivity was
measured using a standard four-probe method. The temperature
dependence of dc magnetization was measured on a Quantum Design
Magnetic Property Measurement System (MPMS-5). The thermopower was
measured by a steady-state technique.

\section{Results and discussion}

The  XRD patterns of BaFe$_{2-x}$Ni$_x$As$_2$ crystals are shown
in Fig.1. Only (00$l$) reflections appear, indicating that the
$c$-axis is perpendicular to the cleaved surface. The (00$l$)
refections with only even $l$ were observed because the system
belongs to the body-centered space group. The $c$-axis lattice
constant was calculated as 1.303 nm for the undoped parent
compound BaFe$_2$As$_2$, consistent with previous
reports\cite{Johrendt-BaFeAs}. The inset of Fig.1 shows the
variation of the $c$-axis lattice constant with nominal Ni content
$x$. The $c$-axis lattice constant decreases monotonically with
increasing $x$. The EDX measurements show that the actual average
Ni content ($x$) is slightly different from the nominal
composition, but the variation of Ni content is less than 5\% in
each sample except the sample with nominal $x$ of 0.05, which
indicates that the distribution of Ni doping is quite uniform in
these samples. A representative EDX spectrum for a single crystal
with nominal $x$ of 0.1 is shown in Fig.2 and the inset shows a
SEM photo of this sample. According to the EDX reports, the
average $x$ value is 0.055, 0.096, 0.18, and 0.23 for the samples
with the nominal $x$ of 0.05, 0.1, 0.16, and 0.2, respectively.

Fig. 3 shows the temperature dependence of in-plane resistivity.
Similar to the previous report\cite{Chen-BaKBaLa,Johrendt-BaFeAs},
there is an anomalous decrease in the resistivity of the parent
compound around $T_{an}$ of 140 K, which is associated with a
structural phase transition. Neutron studies have confirmed that
the spin-density wave order occurs simultaneously with the
structural phase transition in 122
phase\cite{Baowei-BaFeAs,Jesche-SrFeAs,Su-BaFeAs}. The resistivity
anomaly is shifted to a lower temperature, about 90 K, for the $x$
= 0.055 sample. However, the resistivity exhibits a sharp increase
instead of a decrease around $T_{an}$. Such anomalous increase in
the resistivity has been also observed in other doped 122 phase
\cite{SrFeCoAs}. With further decreasing temperature, a
superconducting transition occurs below $T_c^{on}$ (defined as the
onset point in the resistive transition) of 12.5 K for the same
$x$ = 0.055 sample, but zero resistivity can not be reached even
for $T$ as low as 4 K. It is hard to distinguish whether there is
a microscopic phase separation (inhomogeneous distribution of Ni
content) or there is a co-existence of SDW order and
superconducting order according to current measurements. With
increasing Ni content, the resistivity anomaly disappears. At the
optimal doping level $x$ = 0.096, $T_c^{on}$ reaches a maximum of
20.5 K, $T_c^{mid}$ is 20.2 K, and the transition width
$\triangle$$T_c$ is less than 1.0 K, suggesting a sharp
superconducting transition. The temperature dependence of
susceptibility as shown in Fig.2 (b) also indicates very sharp
superconducting transitions for $x$ = 0.096 and 0.18. However the
superconducting transition in magnetic susceptibility of the $x$ =
0.055 sample is rather broad, which means that the distribution of
Ni dopants could be very inhomogeneous in this sample.

Fig.4 shows the variation of $T_c^{on}$ and $T_c^{mid}$ with Ni
content $x$. In the superconducting window (0.055 $\leq$ $x$
$\leq$ 0.23), a dome-like $T_c(x)$ curve is roughly established,
similar to that of Ni doped 1111 phase. It should be noted that
the optimal doping level $x$ = 0.096 corresponds to about 5\% of
Fe substituted by Ni in 122 phase, which is very close to the
optimal Ni doping level in 1111 phase (4\% of Fe substituted by
Ni)\cite{Cao-LaNi}. However, compared to Co-doped 1111 and 122
phases, the superconducting window in Ni-doped system is much
narrower, and the optimal Ni doping concentration (about 5\% Ni
content) is only about half of that of Co-doped
systems\cite{Cao-LaCo, SmCo}. Compared to Fe$^{2+}$ ion, Co$^{2+}$
(3$d^7$) has one more 3$d$ electron, but Ni$^{2+}$ (3$d^8$) has
two more 3$d$ electrons. Then it is expected that each Ni dopant
induces two extra itinerant electrons while each Co dopant only
induces one extra itinerant electron. Thus the fact that the
optimal doing content of Ni is only about half of that of Co can
be understood. Actually both Ni doped and Co doped systems show
maximum $T_c$ at the same effective doping level. A universal
doping (charge carrier concentration) dependence of $T_c$ has been
established in high-$T_c$ superconducting cuprates. Our result
implies that there might exists a universal doping dependence of
$T_c$ in iron-based arsenide superconductors. For the higher
Ni-doping levels ($x\geq$ 0.23), superconductivity is no longer
observed for $T$ $>$ 4 K, but the resistivity becomes even more
metallic. It should be noted that the other end member,
BaNi$_2$As$_2$, is a superconductor with $T_c$ of only about 0.7 K
\cite{Ronning}. However, the first principles calculations
suggested that BaNi$_2$As$_2$ could be a conventional
phonon-mediated superconductor\cite{Singh}.

Fig. 5 plots the thermopower ($S$) as a function of temperature.
All the samples show negative thermopower, which means that
electron-type charge carriers dominate. For the undoped parent
compound ($x$ = 0), thermopower exhibits an anomalous enhancement
just below the structural phase transition temperature $T^{on}$ of
about 140 K. Similar behavior has been observed in undoped parent
compounds such as LaFeAsO\cite{McGuire}, TbFeAsO\cite{ZAXu-TbTh}
and SmFeAsO\cite{SmCo}. With Ni doping, this anomaly is suppressed
quickly. For $x$ = 0.055, only a slight enhancement can be
observed below about 80 K (indicated by the arrow in Fig.4).
Meanwhile the absolute value of normal state thermopower ($|S|$)
increases remarkably with increasing Ni content, and reaches a
maximum at the optimal doping level $x$ = 0.096, then decreases
with further Ni doping. It has been suggested that there is a
correlation between the normal state thermopower and $T_c$
according to the studies on the thermopower of Co-doped SmFeAsO
system\cite{SmCo}. In the Ni-doped 122 phase, such a correlation
should also exist although the absolute value of thermopower is
much smaller than that of F-doped 1111 phase and Co-doped 1111 and
122 phases.

\section{Conclusion}

In conclusion, a series of Ni-doped 122 phase single crystals
BaFe$_{2-x}$Ni$_x$As$_2$ ($x$ = 0, 0.055, 0.096, 0.18, 0.23) were
successfully synthesized by self flux method and superconductivity
with $T_c$ as high as 20 K is observed. The thermopower is
negative in the normal state, suggesting that the electron-like
charge carrier dominates. This Ni-doped system provides another
example of superconductivity induced by electron doping in the 122
phase.

\section*{Acknowledgments}
This work is supported by the National Science Foundation of
China, the National Basic Research Program of China
(No.2006CB601003 and 2007CB925001), and the PCSIRT project of the
Ministry of Education of China (IRT0754)

\pagebreak[4]

\begin{figure}
\caption{\label{Fig1} (Color online) X-ray diffraction pattern at
room temperature for the BaFe$_{2-x}$Ni$_x$As$_2$ single crystals.
The inset shows the variation of the $c$-axis lattice constant
with $x$, where the $x$ values are only the nominal Ni contents
according to starting materials.}
\end{figure}

\begin{figure}
\caption{\label{Fig2} (Color Online) A representative EDX spectrum
for a single crystal with nominal $x$ of 0.1. The inset shows the
SEM photo of this sample. The scale on the bottom of the photo is
100 $\mu$m. The actual Ni content in this sample is among 0.092 to
0.099 according to the EDX spectra.}
\end{figure}

\begin{figure}
\caption{\label{Fig3} (Color Online) (a) Temperature dependence of
in-plane resistivity, and (b) temperature dependence of magnetic
susceptibility for the BaFe$_{2-x}$Ni$_x$As$_2$ single crystals.
The susceptibility was measured under zero-field cooling (ZFC)
condition and the applied magnetic field $H$ is 10 Oe along the
$c$-axis direction. Because of the extremely large demagnetization
factor of the samples the observed superconducting diamagnetic
volume fraction is obviously larger than 100\%.}
\end{figure}

\begin{figure}
\caption{\label{Fig4} (Color Online) The variations of $T_c^{on}$
and $T_c^{mid}$ with Ni content $x$ for BaFe$_{2-x}$Ni$_x$As$_2$
single crystals.}
\end{figure}

\begin{figure}
\caption{\label{Fig5} (Color Online) Temperature dependence of
thermopower ($S$) for BaFe$_{2-x}$Ni$_x$As$_2$ single crystals.
The arrows indicate the anomaly in thermopower associated with
structural phase transition and/or SDW ordering.}
\end{figure}

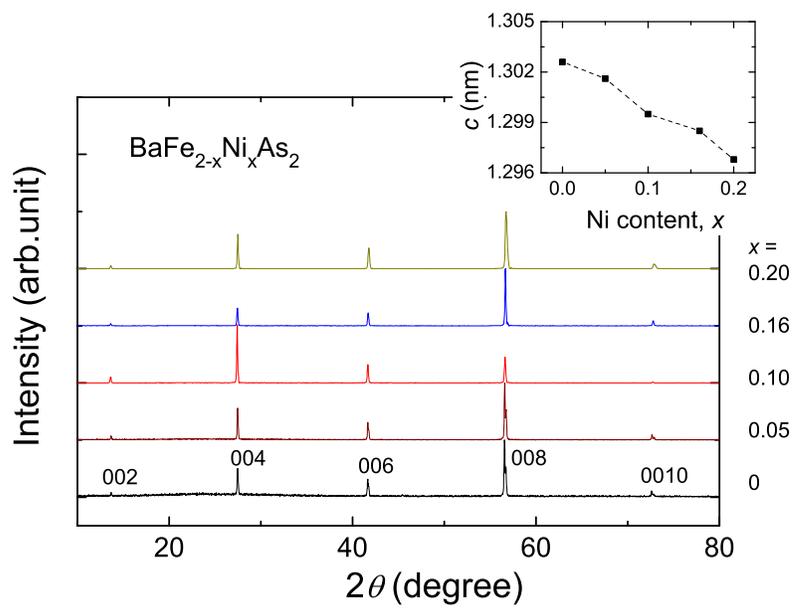
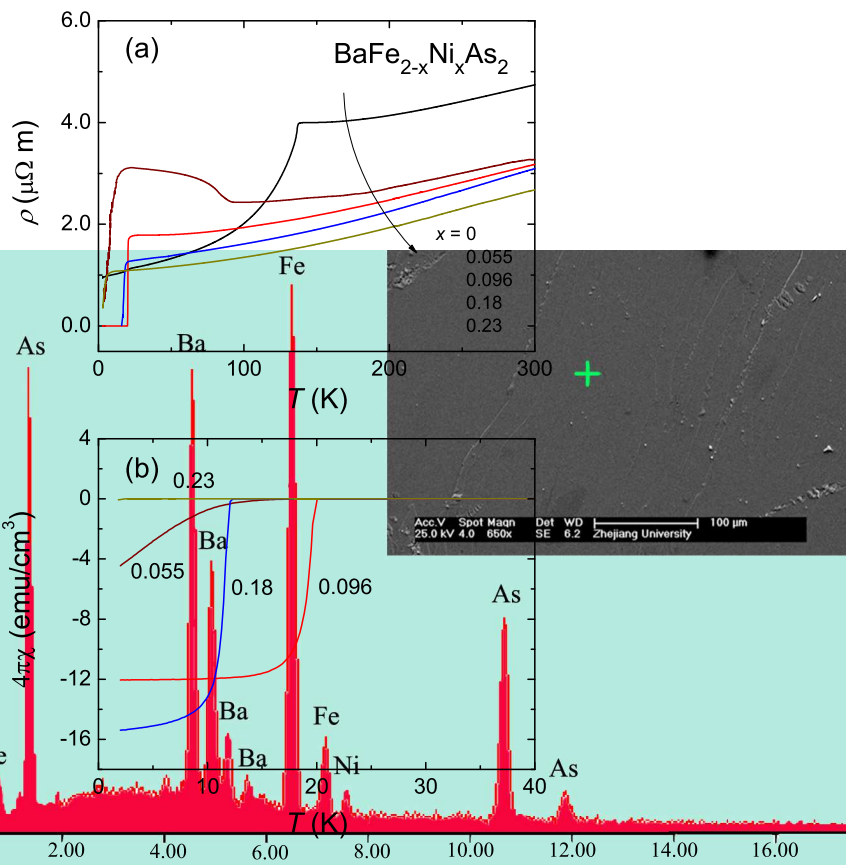
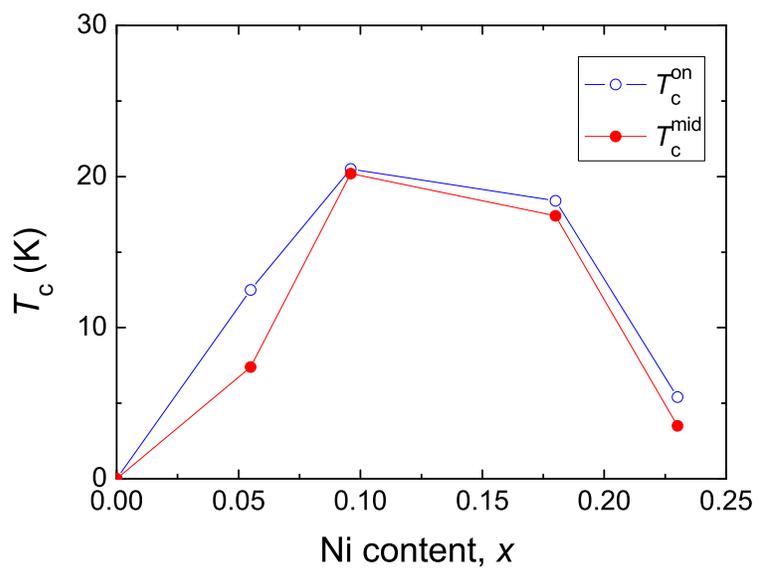
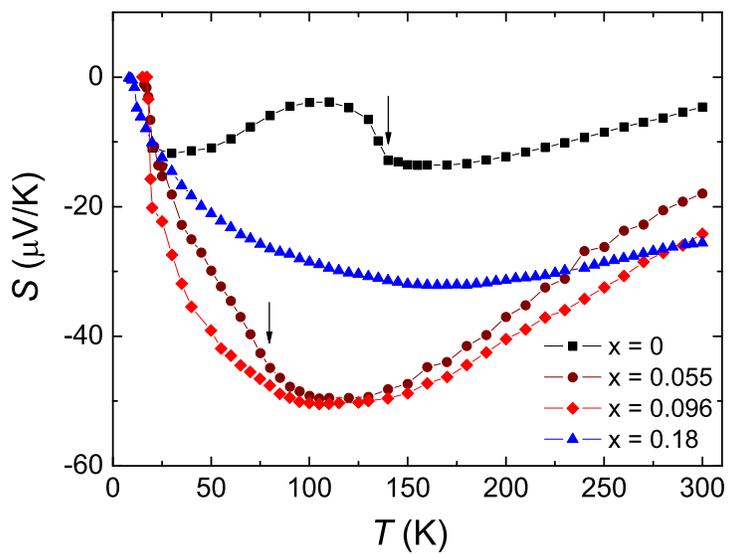
\end{document}